\documentclass{article}
\usepackage{newtxtext}
\usepackage{tikz}
\usepackage{arxiv}
\usepackage{caption}
\usepackage{subcaption}
\usetikzlibrary{positioning, shapes, arrows.meta}
\usepackage{threeparttable}

\usepackage[utf8]{inputenc} 
\usepackage[T1]{fontenc}    
\usepackage{hyperref}       
\usepackage{url}            
\usepackage{booktabs}       
\usepackage{amsfonts}       
\usepackage{nicefrac}       
\usepackage{microtype}      
\usepackage{lipsum}		
\usepackage{graphicx}
\usepackage{array} 
\usepackage{natbib}
\usepackage{doi}
\usepackage{tablefootnote}

\title{SAKURAONE: Empowering Transparent and Open AI Platforms through Private-Sector HPC Investment in Japan}

\author{ \href{https://orcid.org/0009-0002-1241-1379}{\includegraphics[scale=0.06]{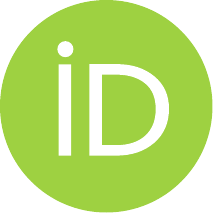}\hspace{1mm}Fumikazu Konishi} \\
	Research Center\\
	SAKURA internet Inc.\\
	Japan\\
	\texttt{f-konishi@sakura.ad.jp} \\
}

\renewcommand{\headeright}{}
\renewcommand{\undertitle}{}
\renewcommand{\shorttitle}{SAKURAONE}

\hypersetup{
pdftitle={SAKURAONE: Empowering Transparent and Open AI Platforms through Private-Sector HPC Investment in Japan},
pdfsubject={q-bio.NC, q-bio.QM},
pdfauthor={Fumikazu KONISHI},
pdfkeywords={First keyword, Second keyword, More},
}

\begin{document}
\maketitle

\begin{abstract}
SAKURAONE is a managed high performance computing (HPC) cluster developed and operated by the SAKURA Internet Research Center. It reinforces the ``KOKARYOKU PHY'' configuration of bare-metal GPU servers and is designed as a cluster computing resource optimized for advanced workloads, including large language model (LLM) training.

In the ISC 2025 edition of the TOP500 list, SAKURAONE was ranked \textbf{49th} in the world based on its High Performance Linpack (HPL) score, demonstrating its global competitiveness. In particular, it is the \textbf{only system within the top 100} that employs a fully open networking stack based on \textbf{800~GbE (Gigabit Ethernet)} and the \textbf{SONiC (Software for Open Networking in the Cloud)} operating system, highlighting the viability of open and vendor-neutral technologies in large-scale HPC infrastructure.

SAKURAONE achieved a sustained performance of 33.95~PFLOP/s on the HPL benchmark (Rmax), and 396.295~TFLOP/s on the High Performance Conjugate Gradient (HPCG) benchmark. For the HPL-MxP benchmark, which targets low-precision workloads representative of AI applications, SAKURAONE delivered an impressive 339.86~PFLOP/s using FP8 precision.

The system comprises 100 compute nodes, each equipped with eight NVIDIA H100 GPUs. It is supported by an all-flash Lustre storage subsystem with a total physical capacity of 2~petabytes, providing high-throughput and low-latency data access.
Internode communication is enabled by a full-bisection bandwidth interconnect based on a Rail-Optimized topology, where the Leaf and Spine layers are interconnected via 800~GbE links. This topology, in combination with RoCEv2 (RDMA over Converged Ethernet version 2), enables high-speed, lossless data transfers and mitigates communication bottlenecks in large-scale parallel workloads.
\end{abstract}



\section{Introduction}

SAKURA internet inc. has focused on server and cloud infrastructure services since its founding in the early days of the Internet. In recent years, cloud computing has become increasingly essential as a core part of modern societal infrastructure, and its role is expected to grow further in the development of advanced technologies and industries.

In the United States, major technology companies, often referred to as “Big Tech”—have established in-house AI infrastructures, enabling them to conduct cutting-edge research and development in artificial intelligence. In contrast, while some Japanese companies are developing their own AI applications, they remain relatively small in scale and lack the robust computational infrastructure necessary to support such efforts. There is a growing need for private sector entities in Japan capable of making large-scale, end-to-end investments in AI infrastructure comparable to their international counterparts.

Within Japan, \cite{takano2024abci30evolutionleading} and \cite{10.1007/978-3-031-73716-9_16} shared high-performance computing services such as ABCI3.0, operated by the National Institute of Advanced Industrial Science and Technology (AIST), and TSUBAME4.0 , provided by the Tokyo Institute of Science, have been made available to industrial users. However, these systems are typically shared with academic users, making it difficult for private sector organizations to secure computing resources in a stable and predictable manner. As such, there is an increasing demand for equally capable systems that can be provided by the private sector and dedicated to consistent commercial use.

In response to this situation, SAKURA launched a large-scale investment initiative in 2023, committing approximately 13 billion JPY toward the expansion of its cloud infrastructure, including the deployment of 2,000 NVIDIA GPUs. In 2024, this initiative was further accelerated with governmental support for an additional investment project valued at approximately 100 billion JPY, aimed at establishing an even more extensive AI computing foundation.

By significantly strengthening computational resources for AI, SAKURA seeks not only to contribute to the advancement of artificial intelligence technologies, but also to support the broader development of next-generation cloud infrastructure, including Japan's government cloud platforms. The SAKURAONE system has been designed as a cornerstone of these efforts, providing a high-performance computing platform capable of meeting the growing demands of AI research, development, and industrial applications within Japan.

\begin{figure}
	\centering
        \includegraphics[width=0.8\linewidth]{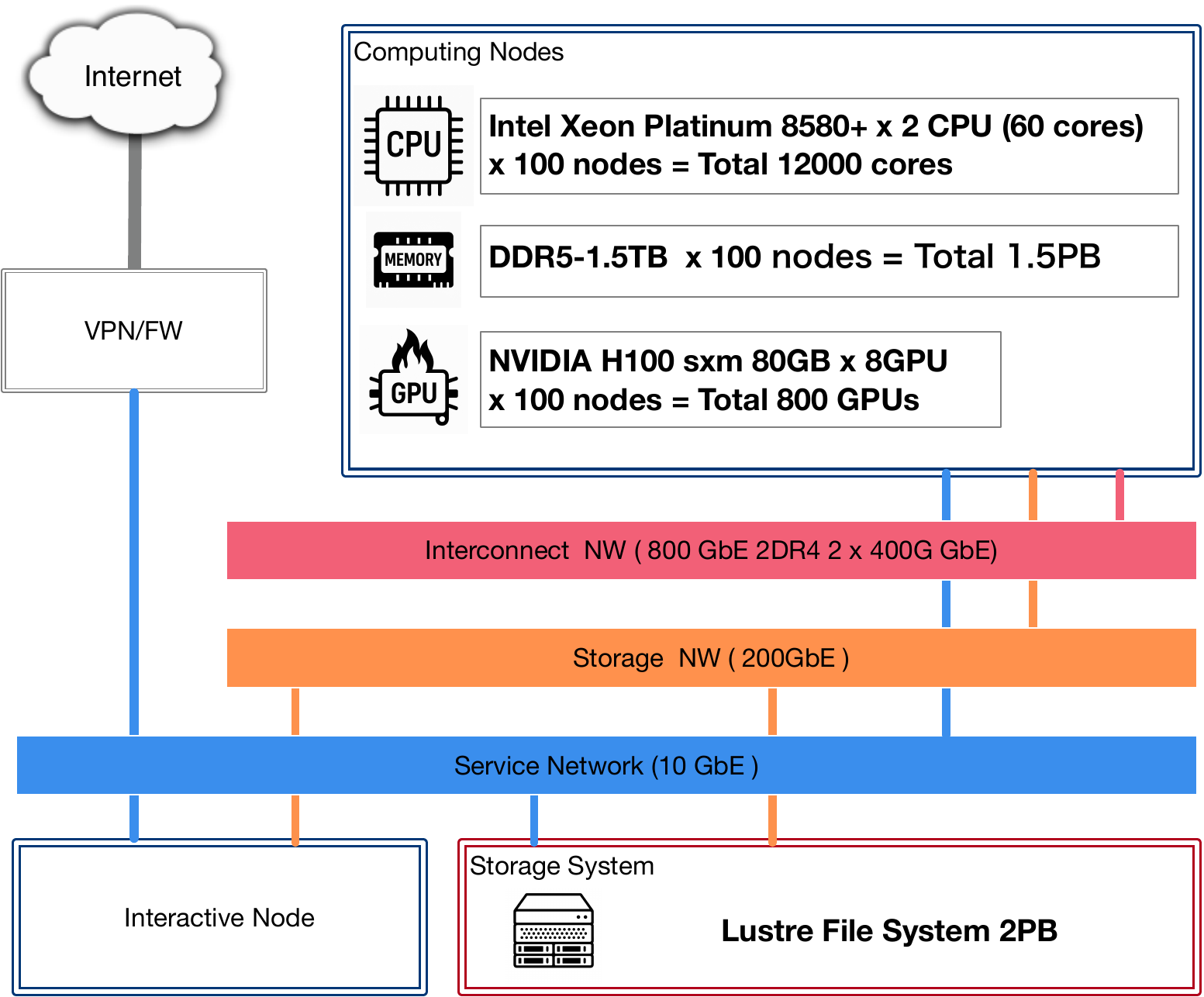}  
	\caption{SAKURAONE System Overview}
	\label{fig:system_overview}
\end{figure}

Figure \ref{fig:system_overview} shows an overview of SAKURAONE. It comprises 100 compute nodes, each equipped with eight NVIDIA H100 GPUs, totaling 800 GPUs across the cluster. The system includes a two-petabyte all-flash Luster storage solution, which provides high-throughput and low-latency data access. GPU-to-GPU communication is enabled by a full-bisection interconnect configured in a rail-optimized topology and connected via RoCEv2, allowing high-speed data transfer between nodes. In addition, access to interactive front-end nodes is secured and accelerated via a high-speed VPN connection, facilitating efficient remote use.

\section{SAKURAONE Hardware}
The hardware configuration of SAKURAONE comprises computational nodes, which handle primary computing tasks; high-speed interconnect switches, facilitating efficient communication between nodes; and a storage system based on a high-performance parallel filesystem, designed to manage intensive I/O demands from these compute nodes. To ensure stability of performance across the entire system, the interconnect network and the storage network are logically separated, minimizing interference between computational and storage traffic. In addition, dedicated interactive nodes are provided separately, enabling users to submit and manage computational jobs efficiently. In addition, a VPN gateway has been implemented to isolate SAKURAONE from direct internet access, establishing a secure and robust operational environment.

\subsection{Computing Node}
The SAKURAONE computing node, summarized in Table~\ref{tab:compute_node}, is based on the Supermicro GPU SuperServer SYS-821GE-TNHR, an air-cooled 8U platform optimized for high-performance computing and AI workloads. Each node is equipped with two Intel Xeon Platinum 8580+ processors (5th Gen, 60 cores, 120 threads per CPU), totaling 120 cores per node. The CPUs support eight-channel DDR5-5600 memory, with a total capacity of 1.5~TB, enabling high-throughput memory access for large-scale parallel applications.

The system integrates eight NVIDIA H100 SXM 80GB GPUs connected via NVLink and NVSwitch, providing high-bandwidth GPU-to-GPU communication essential for distributed deep learning. Local storage is provisioned via four 7.68~TB NVMe SSDs and dual 372~GB SAS drives. Networking includes eight 400~GbE/NDR interfaces for inter-node communication, two 400~GbE/NDR interfaces for storage traffic, and additional low-speed interfaces for management.

To analyze GPU-to-NIC connectivity, we utilized the \texttt{nvidia-smi topo -mp} command. Based on this output, Table~\ref{tab:nic-usage} summarizes the classification and types of PCIe connectivity for each NIC relative to GPUs.
NIC0–NIC7 are directly associated with GPU0–GPU7 through \texttt{NODE}-level PCIe paths, providing NUMA-local, low-latency connectivity ideal for high-throughput collective operations via NCCL or MPI over RoCEv2. NIC8 and NIC10 are connected via longer \texttt{PXB} paths and are inferred to serve the storage network. NIC10 represents a bonded interface (\texttt{mlx5\_bond\_0}), aggregating multiple physical links for redundancy or increased bandwidth. NIC9 is connected via a \texttt{SYS}-level path that crosses NUMA boundaries and is presumed to support management plane traffic such as SSH.

This hardware topology reflects a deliberate design strategy that separates compute, storage, and control traffic at the hardware level. Such partitioning enhances communication efficiency and fault isolation, which are critical for scalable and reliable AI training workloads.

\begin{table}
	\caption{SAKURAONE Computing Nodes}
	\centering
	\begin{tabular}{ll}
		\toprule
		Name     & Description      \\
		\midrule
            \midrule    
            Chassis & Supermicro GPU SuperServer SYS-821GE-TNHR      \\
		CPU & Intel Xeon Platinum 8580+ x 2 CPUs      \\
		Core (per CPU) & 120 (60)     \\        
		GPU     & NVIDIA H100 SXM 80GB x 8 GPUs      \\
		Memory (RAM)     & DDR5-1.5TB (96GB x 16)         \\
		System storage (SAS)    & 372GB x 2        \\
		  Data storage (NVMe)    & 7.68TB x 4         \\
		  Local network interface     & 4GbE (2Gbps x 2)         \\
		Interconnet network interface & NVIDIA Mellanox MCX75310AAS-NEAT ConnectX®-7 400GbE/NDR x 8         \\
		Storage network interface    & NVIDIA Mellanox MCX75310AAS-NEAT ConnectX®-7 400GbE/NDR x 2        \\
		\bottomrule
	\end{tabular}
	\label{tab:compute_node}
\end{table}

\begin{table}[htbp]
\centering
\begin{threeparttable}
\caption{Classification of NIC Usage and GPU Connectivity Characteristics}
\label{tab:nic-usage}
\begin{tabular}{clll}
\toprule
\textbf{NIC} & \textbf{Device Name} & \textbf{Primary Usage} & \textbf{GPU Connectivity Type} \\
\midrule
NIC0  & mlx5\_0        & High-speed inter-node communication     & NODE (via GPU0 PCIe domain) \\
NIC1  & mlx5\_1        & High-speed inter-node communication     & NODE (via GPU1 PCIe domain) \\
NIC2  & mlx5\_2        & High-speed inter-node communication     & NODE (via GPU2 PCIe domain) \\
NIC3  & mlx5\_3        & High-speed inter-node communication     & NODE (via GPU3 PCIe domain) \\
NIC4  & mlx5\_4        & High-speed inter-node communication     & NODE (via GPU4 PCIe domain) \\
NIC5  & mlx5\_5        & High-speed inter-node communication     & NODE (via GPU5 PCIe domain) \\
NIC6  & mlx5\_6        & High-speed inter-node communication     & NODE (via GPU6 PCIe domain) \\
NIC7  & mlx5\_7        & High-speed inter-node communication     & NODE (via GPU7 PCIe domain) \\
\midrule
NIC8  & mlx5\_8        & Storage network (dedicated I/O path)    & PXB \\
NIC10 & mlx5\_bond\_0  & Storage network (bonded for redundancy) & PXB (logical, multi-bridge path) \\
\midrule
NIC9  & mlx5\_11       & Management network (e.g., SSH)     & SYS  \\
\bottomrule
\end{tabular}
\begin{tablenotes}
\footnotesize
\item This classification is based on \texttt{nvidia-smi topo -mp} output, analyzing PCIe connectivity and NUMA affinity between GPUs and NICs. NIC0–NIC7 are directly mapped to GPUs 0–7 using NODE-level PCIe paths for optimized inter-node communication. NIC8 and NIC10 are likely dedicated to storage traffic and are accessed via longer PXB paths or logical bonding. NIC9 is connected to the system via a SYS-level path that crosses NUMA domains.
\end{tablenotes}
\end{threeparttable}
\end{table}

\subsection{Network System}

Continuous advancements in high-performance computing (HPC) have underscored the critical importance of high-performance interconnect networks, which significantly influence the efficiency and scalability of HPC applications. In designing the SAKURAONE interconnect architecture, we carefully considered the recent trends observed among the systems listed in the TOP500 rankings.

Interconnect networks are typically evaluated based on several key parameters, including bandwidth, latency, switch radix, and overall network topology. According to a 2022 analysis of the TOP500 list, InfiniBand was the most widely adopted interconnect technology, dominating the majority of high-performance system deployments. In contrast, only one system within the top 10 utilized gigabit Ethernet (GbE) at that time. However, a more recent review of the November 2024 TOP500 list indicates a notable shift: Seven out of the top ten systems now employ GbE-based interconnects.

A major factor contributing to this trend is the increased adoption and maturity of Remote Direct Memory Access (RDMA) technologies in Ethernet-based HPC environments. In particular, RoCEv2 (RDMA over Converged Ethernet version 2) has emerged as a viable high-performance alternative to InfiniBand, delivering competitive performance over standard Ethernet infrastructure.
For SAKURAONE, we selected GbE with RoCEv2 as the interconnect technology due to its lower deployment and operational costs, as well as the ability to leverage existing expertise and infrastructure within our data center environment.
Moreover, recent evaluations (e.g. \cite{10820803}) have shown that, aside from latency, there is no significant difference in overall interconnect performance between InfiniBand and RoCEv2. This finding further supports our decision to adopt GbE as a cost-effective and technically robust solution for HPC-scale workloads.

\begin{table}[htbp]
\centering
\caption{Interconnect Usage (2020--2024) in Top 10 of Top500 list on Nov 2024 }
\label{tab:interconnect_usage}
\begin{tabular}{lcccccc}
\hline
 & \textbf{2020} & \textbf{2021} &  \textbf{2022} &\textbf{2023} & \textbf{2024} & \textbf{Total} \\ 
 \hline
 \hline
\textbf{Gigabit Ethernet} & & 1 & & 2 & 4 & 7 \\
\hline
\hspace{5mm}Slingshot-11 & & 1 & & 2 & 4 & 7 \\
\hline
\textbf{Infiniband} & & & & 2 & & 2 \\
\hline
\hspace{5mm}NVIDIA Infiniband NDR & & & & 1 & & 1 \\
\hline
\hspace{5mm}Quad-rail NVIDIA HDR100 Infiniband & & & & 1 & & 1 \\
\hline
\textbf{Proprietary Network} & 1 & & & & & 1 \\
\hline
\hspace{5mm}Tofu interconnect D & 1 & & & & & 1 \\ \hline
\hline
\textbf{Total} & 1 & 1 & & 4 & 4 & 10 \\ \hline
\end{tabular}
\end{table}

In modern high-performance computing (HPC) systems tailored for AI workloads, the design of the interconnect topology plays a pivotal role in determining training efficiency and scalability. To address the communication demands of large-scale distributed AI systems, established and emerging network topologies have been adopted. In the following, we outline four representative topological configurations commonly observed in AI clusters on a production scale or in a research oriented environment, as summarized in~\cite{duan2024efficienttraininglargelanguage}.

\paragraph{Fat-Tree Topology}
Fat-tree topologies, derived from Clos network theory, are widely deployed in traditional HPC environments. They provide full bisection bandwidth through a hierarchical switch structure, offering balanced communication paths and minimizing network congestion. Fat-tree networks are particularly well suited for workloads that involve frequent all-to-all communication, such as data-parallel deep learning.

\paragraph{Dragonfly Topology}
Dragonfly networks organize compute nodes into groups, where each group typically consists of multiple switches (e.g., leaf switches) that are fully connected to one another and to a subset of compute nodes. Within a group, nodes can communicate through high-bandwidth, low-latency intragroup links, enabling efficient local communication.

Each group also maintains a limited number of global links that connect to switches in other groups. This hierarchical yet flattened structure reduces the number of hops required for global communication and ensures high global bandwidth with fewer long-distance connections compared to traditional topologies like torus or Fat-Tree topology.

The group in this context can be understood as a logical and often physical unit, such as a rack or a set of adjacent racks, designed to optimize both intragroup data movement and intergroup scalability. In practical deployments, this topology is particularly effective for large-scale AI workloads characterized by intensive collective communication patterns (e.g., all-reduce in distributed training), where minimizing latency and maximizing bisection bandwidth are critical to performance.

\paragraph{Rail-Only Topology }
As reported in \cite{Wang_2024}, rail-only topologies connect GPUs or accelerators via a dedicated, flat, high-bandwidth network that excludes shared storage or external I/O paths. This configuration isolates compute traffic and reduces contention, making it ideal for tightly coupled training jobs that require synchronized GPU communication across nodes.

\paragraph{Rail-Optimized Topology}
Extending the concept of rail only, rail-optimized topologies incorporate features such as redundant paths, adaptive routing, and hierarchical interconnect layers to improve scalability, bandwidth efficiency, and fault tolerance. These designs are prevalent in GPU-dense supercomputers where the interconnect must scale proportionally with node count while sustaining high performance under intensive communication loads.

In designing the network topology for our GPU cluster, we evaluated several reference architectures and ultimately selected a configuration optimized for shortest-path communication between GPUs. Although Fat-Tree and Dragonfly topologies offer well-established advantages in traditional HPC environments and theoretical performance models, Fat-Tree topology was deemed suboptimal for AI-centric workloads at scale.
Although Dragonfly provides promising performance characteristics, we identified a lack of in-house technical expertise necessary to fully design, fine-tune, and operate such a topology within our deployment environment.
Taking into account the demonstrated effectiveness of rail-only designs in large language model (LLM) training, along with the enhanced scalability and resilience provided by rail-optimized configurations, we adopted a rail-optimized topology for our system.

Each of these topologies addresses different challenges in scaling AI workloads—from minimizing inter-node latency to maximizing aggregate bandwidth and system utilization. Our selection process was guided by the specific performance requirements of the target LLM applications, the anticipated scale of the cluster, and the infrastructure cost constraints. The resulting topology represents a balanced solution that emphasizes performance, flexibility, and cost efficiency.

To further reduce deployment costs while preserving high network performance, we implemented a white-box switching architecture. The software stack is based on SONiC (Software for Open Networking in the Cloud), and the hardware is powered by Broadcom’s Tomahawk 5 switching ASIC, providing state-of-the-art forwarding capabilities and support for high-radix configurations.

\begin{figure}
	\centering
         \includegraphics[width=0.8\linewidth]{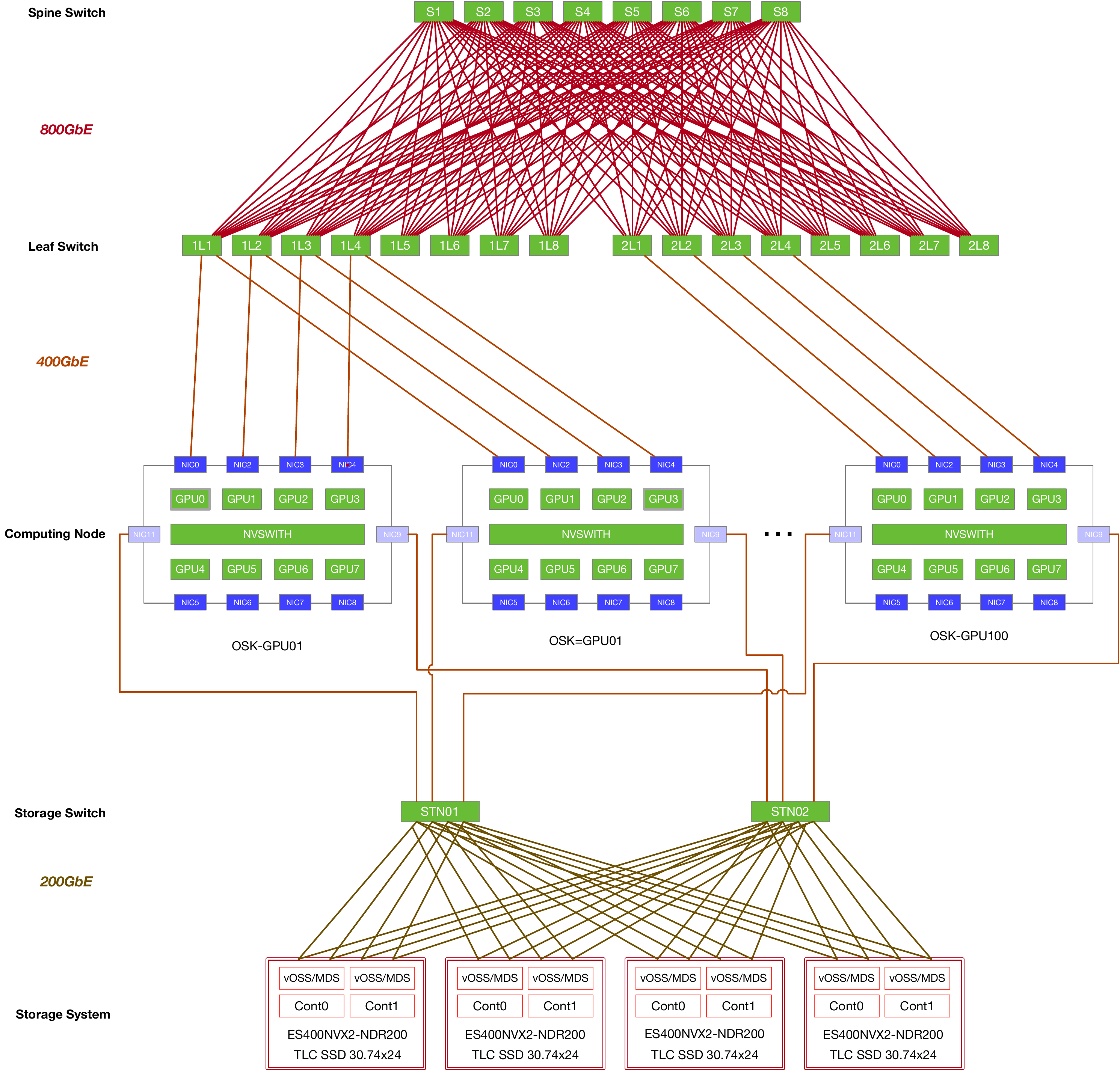}  
	\caption{SAKURAONE Network Overview}
	\label{fig:network_overview}
\end{figure}

Figure \ref{fig:network_overview} illustrates the schematic connectivity between the computing nodes in SAKURAONE and the Leaf and Spine switches that form the Rail-Optimized network topology. The system’s 100 computing nodes are divided into two groups of 50 nodes each, designated as Pod 1 and Pod 2. Each computing node is equipped with eight GPUs, which are connected via dedicated 400 GbE network links to a corresponding set of eight Leaf switches within each pod.

The 16 Leaf switches across both pods are connected to a shared set of eight Spine switches, forming a high-bandwidth, scalable Rail-Optimized topology. Each leaf switch is directly linked to every spine switch via 800 GbE interconnects, enabling uniform, high-throughput communication across the cluster and minimizing latency for bandwidth-intensive AI workloads.

Table \ref{tab:intercoonect_network} summarizes the specifications of the interconnect network configuration used in SAKURAONE.

\begin{table}
	\caption{SAKURAONE Interconnect Network}
	\centering
	\begin{tabular}{ll}
		\toprule
		Name     & Description      \\
		\midrule
            Network technology & Gigabit Ethernet (GbE)     \\
		  Ethernet switch speed grdade & 800 GbE 2DR4 2 x 400G GbE    \\
		Protocol & RoCEv2 (RDMA over Converged Ethernet)
     \\        
		Network topology     & Rail-Optimized Topology      \\
		  Switch Chassis & Edge-core networks AIS800-64O \\
            Switch Capability & 51.2Tbps fullduplex \\
		Software Stack & SONiC (Software for Open Networking in the Cloud)      \\
		  Switch Chip    & Tomahawk 5 chipset powered by Broadcom        \\
		  Leaf Switch     & 16 chassis\tablefootnote{The 16 Leaf switches are organized into two pods of 8 switches each, with each pod forming a dedicated rail fabric connected to a distinct set of compute nodes.}         \\
		Spine Switch & 8  chassis      \\
		\bottomrule
	\end{tabular}
	\label{tab:intercoonect_network}
\end{table}

\subsection{Storage system}

SAKURAONE is equipped with a three-tier storage architecture designed to meet the diverse performance and capacity requirements of large-scale computing workloads.

Each compute node is provisioned with a local storage subsystem dedicated to system operations, including the operating system. This comprises a 372 GB RAID1 volume utilizing mirrored drives for redundancy and reliability. In addition to the system volume, each node is equipped with four 7.68 TB NVMe drives, providing 30.72 TB of high-speed local storage. This local storage is primarily intended for temporary data during computation and scratch use at the node level.
For large-scale workloads that demand high-throughput parallel I/O, SAKURAONE deploys a high-performance Lustre file system shared across all 100 compute nodes. This shared file system offers a total capacity of 2 PB and is used to store checkpoint data and intermediate results during computational tasks such as training of large language models (LLM).

Each compute node is connected to the storage network via two 400 GbE interfaces, ensuring high-bandwidth access with network redundancy. The backend of the Lustre file system is supported by four DDN ES400NVX2-NDR200 storage servers, as summarized in Table~\ref{tab:storage_system}.”
Each server is equipped with dual active controllers, running vOSS and MDS services, respectively, enabling robust failover and parallel file system operations. The servers feature eight 200 GbE interfaces each, redundantly connected to two storage switches to achieve load-balanced and fault-tolerant connectivity. In the event of a single switch failure, although the aggregate bandwidth is reduced by half, the storage system continues to operate without service interruption.
The theoretical maximum I/O throughput of the shared Lustre file system is 200 GB/s for both read and write operations.

\begin{table}
	\caption{SAKURAONE Storage System}
        \label{tab:storage_system}
	\centering
	\begin{tabular}{lll}
		\toprule
		Name     & Description  & Count  \\
		\midrule
            Chassis    & DDN ES400NVX2  & 4       \\
          \bottomrule
            Controller    & Active Dual Controller & 2  \\
            CPU    & Ice Lake CPUs  & 2   \\
            NVMe    & 24 Drive (PCI Gen4 ) & 24   \\        
		Drive  & TLC SSD 30.72TB    &   \\
            Interface  & 200 GbE &  4      \\
		\bottomrule
	\end{tabular}

\end{table}

\section{SAKURAONE System Software}
The target system in this study is SAKURAONE, a high-performance computing (HPC) cluster designed to support a wide range of scientific and engineering workloads. The system is built on Rocky Linux, an enterprise-grade Linux distribution that is community-driven and fully compatible with Red Hat Enterprise Linux (RHEL). The adoption of Rocky Linux ensures long-term stability, reliable security updates, and consistent compatibility with a wide array of HPC-optimized software packages. These characteristics make it particularly well suited for production HPC environments, where software reliability, reproducibility, and integration with established scientific toolchains are critical.

In modern HPC, cluster systems are expected not only to support high-precision numerical computation and high-throughput data transfer, but also to provide an integrated software ecosystem capable of supporting increasingly complex research and development workflows. SAKURAONE addresses these requirements by offering a comprehensive software environment that includes state-of-the-art middleware, parallel programming libraries, and development tools widely used across computational science, physics, and machine learning communities.

The system provides full support for conventional parallel programming paradigms such as MPI and OpenMP, along with development environments tailored for recent GPU and AI accelerator architectures. Development kits such as CUDA and compilers with GPU offloading capabilities enable efficient implementation of compute-intensive tasks, both for traditional simulations and deep learning workloads.Deep learning frameworks such as TensorFlow and PyTorch, as well as GPU-optimized libraries including cuDNN and TensorRT, are available in a modular and user-friendly format. This allows researchers to quickly deploy and use powerful AI toolchains without the burden of managing complex dependencies or conflicting environments. The availability of such standardized and modular environments significantly reduces the barrier to entry for novice users, while also supporting reproducible workflows for expert practitioners.

To further improve software portability and reproducibility, SAKURAONE supports container-based virtualization technologies, including Singularity(Apptainer) and Pyxis, which are widely used in HPC contexts. These container solutions allow users to encapsulate software stacks and dependencies into lightweight, immutable environments that can be deployed consistently across different nodes or even different systems. As a result, source code and runtime configurations require minimal modification when scaling applications from development to large-scale execution.

To ensure efficient and fair resource utilization across a multi-user, multi-project environment, the SAKURAONE cluster incorporates the Slurm workload manager. Slurm is an open-source, scalable, and highly configurable job scheduler commonly used in top-tier supercomputers. It enables users to submit and manage jobs in a controlled and policy-compliant manner, supporting advanced features such as job prioritization, node reservation, resource limits, and job dependencies. These capabilities are essential for handling diverse workloads with varying resource requirements and execution patterns.

In addition, Slurm integration with performance monitoring and user-level job control tools allows administrators and researchers to gain real-time insight into resource usage, diagnose performance bottlenecks, and fine-tune scheduling strategies. In this way, the system maintains a high level of operational efficiency and supports sustained high-throughput execution even under heavy-use scenarios.

In summary, SAKURAONE combines a stable and HPC-optimized operating system (Rocky Linux), a flexible and scalable job scheduler (Slurm), and a comprehensive stack of development and runtime tools to deliver a software environment optimized for both traditional high-performance computing and modern AI workloads. This environment supports demanding scientific applications such as structural analysis, testbed simulations, and deep learning-based research, while enabling rapid, reliable, and reproducible development and execution workflows for users at all levels of expertise.

\begin{table}
	\caption{SAKURAONE System Software}
	\centering
        \begin{tabular}{ll}
        \toprule
        Usage & Description \\
        \midrule
        \midrule
        OS    & Rocky Linux release 9.4 (Blue Onyx)     \\
        \midrule
        Container & singularity-ce version 4.3.1-1.el9 \\
        \midrule
        Job scheduler    & slurm 22.05.9  \\
        \midrule
        GPU programming environment & cuda/12.1 \\
         & cuda/12.2 \\
         & cuda/12.4 \\
         & cuda/12.5 \\
         & cuda/12.6 \\
         & cuda/12.8 \\
         \bottomrule
        Deep learning acceleration library & cudnn/8.9.7 \\
         & cudnn/9.4.0 \\
         & cudnn/9.6.0 \\
         \bottomrule
        MPI and communication middleware & hpcx/2.17.1-gcc-cuda12/hpcx \\
         & hpcx/2.17.1-gcc-cuda12/hpcx-debug \\
         & hpcx/2.17.1-gcc-cuda12/hpcx-mt \\
         & hpcx/2.17.1-gcc-cuda12/hpcx-prof \\
         & hpcx/2.17.1-gcc-cuda12/hpcx-stack \\
         & hpcx/2.18.1-gcc-cuda12/hpcx \\
         & hpcx/2.18.1-gcc-cuda12/hpcx-debug \\
         & hpcx/2.18.1-gcc-cuda12/hpcx-mt \\
         & hpcx/2.18.1-gcc-cuda12/hpcx-prof \\
         & hpcx/2.18.1-gcc-cuda12/hpcx-stack \\
        \bottomrule   
        Python environment with ML frameworks & miniconda/24.7.1-py311 \\
         & miniconda/24.7.1-py311-pytorch \\
         & miniconda/24.7.1-py311-tensorflow \\
         & miniconda/24.7.1-py312 \\
         & miniconda/24.7.1-py312-pytorch \\
         & miniconda/24.7.1-py312-tensorflow \\
        \bottomrule   
        GPU collective communication library & nccl/2.20.5 \\
         & nccl/2.21.5 \\
         & nccl/2.22.3 \\
         & nccl/2.23.4 \\
         & nccl/2.24.3 \\
        \bottomrule
        \end{tabular}
	\label{tab:system_software}
\end{table}

\section{Performance Evaluation}
To enable a comparative evaluation of the SAKURAONE system against other general-purpose HPC clusters, we performed a benchmark-based performance analysis using a set of widely recognized tools. These benchmarks provide a standardized means of assessing the computational and communication capabilities of the system in relation to established systems in the field.

Each selected benchmark has been developed with distinct objectives and workload characteristics. Therefore, we first summarize the purpose and design philosophy of each benchmark, followed by a presentation of the corresponding measurement results. This approach enables a multifaceted characterization of SAKURAONE performance in a variety of HPC-relevant metrics.

\paragraph{High Performance Linpack (HPL)}
The High Performance Linpack (HPL) benchmark forms the basis of the TOP500 rankings of the world's most powerful supercomputers. Developed by Jack Dongarra and colleagues, HPL solves dense systems of linear equations using double-precision arithmetic and remains a standard metric for evaluating sustained computational performance. Since its introduction in 1993, HPL has served as a key indicator of raw processing power and system stability under continuous load~\cite{Dongarra2001}.

Table~\ref{tab:hpl_summary} summarizes the results of our HPL benchmark execution. The benchmark was used with the HPL-NVIDIA 25.4.0 implementation. The benchmark used a matrix of size $N = 2,\!706,\!432$ with a block size of $NB = 1024$, distributed over a $16 \times 49$ process grid, totaling 784 processes (and GPUs). The system achieved a sustained performance of 33.95 PFLOPS, which corresponds to 43.31 TFLOPS per GPU. The maximum performance of GEMM on a single GPU was measured at 55.34 TFLOPS. Each GPU was equipped with 132 streaming multiprocessors and operated at a peak clock frequency of 1980 MHz. The entire benchmark was completed in 389.23 seconds, demonstrating the system's high computational efficiency and scalability.

\begin{table}[htbp]
\centering
\caption{HPL Benchmark Summary}
\begin{tabular}{ll}
\toprule
\textbf{Item} & \textbf{Value} \\
\toprule
Matrix size (N) & 2,706,432 \\
Block size (NB) & 1024 \\
Process grid (P×Q) & 16 × 49 \\
Total processes & 784 \\
Total GPUs  & 784 \\
\midrule
HPL version & HPL-NVIDIA 25.4.0 \\
Execution time (sec) & 389.23 \\
FLOPS & 33.95 PFLOPS \\
FLOPS per GPU & 43.31 TFLOPS \\
Max GEMM performance (single GPU) & 55.34 TFLOPS \\
GPU SM count & 132 \\
GPU peak clock frequency & 1980 MHz \\
\bottomrule
\end{tabular}
\label{tab:hpl_summary}
\end{table}

\paragraph{High Performance Conjugate Gradients (HPCG)}
Although HPL effectively assesses floating-point throughput, it lacks relevance for memory- and communication-intensive applications. To address this, \cite{osti_1089988} proposed the High Performance Conjugate Gradients (HPCG) benchmark, which models sparse matrix computations using the iterative conjugate gradient method. Introduced in 2014, HPCG emphasizes performance in realistic workloads, including irregular memory access and global communication, offering a complementary view to HPL.

Table~\ref{tab:hpcg_summary} presents a summary of the HPCG benchmark results. The benchmark was performed using HPCG version 3.1 with a total of 784 distributed processes, each using 16 threads. The global problem dimensions were $4096 \times 3584 \times 3808$, resulting in a system of approximately 55.9 billion equations and 1.51 trillion nonzero terms. The benchmark consumed approximately 39.96 TB of memory in total, of which 35.17 TB were allocated for the linear system and the conjugate gradient solver. An observed peak memory bandwidth of 3.316 TB/s was achieved. The raw computational throughput reached 437.4 GFLOP/s, and after accounting for convergence overhead, the performance was reduced to 404.96 TFLOP/s. The final validated HPCG result was 396.3 TFLOP / s, highlighting the system's ability to handle workloads containing memory and intensive communication efficiently.

\begin{table}[htbp]
\centering
\caption{HPCG Benchmark Summary}
\begin{tabular}{ll}
\toprule
\textbf{Item} & \textbf{Value} \\
\toprule
Benchmark version & HPCG 3.1 \\
Total distributed processes & 784 \\
Threads per process & 16 \\
Global problem dimensions (nx × ny × nz) & 4096 × 3584 × 3808 \\
Number of equations & 55.9 billion \\
Number of nonzero terms & 1.51 trillion \\
Total memory used (GB) & 39,961.4 \\
Memory for linear system and CG (GB) & 35,169 \\
Peak memory bandwidth (observed) & 3.316 TB/s \\
Total GFLOP/s (raw) & 437,361 \\
GFLOP/s (with convergence overhead) & 404,964 \\
Final validated HPCG GFLOP/s result & 396,295 \\
\bottomrule
\end{tabular}
\label{tab:hpcg_summary}
\end{table}

\paragraph{HPL-MxP (AI)}
Traditional benchmarks such as HPL and HPCG focus on double-precision performance, which is less representative of AI workloads. Modern AI training relies heavily on low-precision arithmetic, which benefits from GPU architectures optimized for high-throughput, reduced-accuracy computations. To better reflect this, the HPL-AI (High Performance Linpack for AI) benchmark introduces mixed-precision computation with iterative refinement to ensure numerical accuracy. This makes it a more realistic performance metric for evaluating AI-capable HPC systems~\cite{10.1109/SC.2018.00050}.

Table~\ref{tab:hpl_mxp_summary} summarizes the results of our HPL-MxP benchmark using HPL-MxP-NVIDIA version 25.4.0. The test used a matrix of size $N = 2,\!989,\!056$ with a block size of $NB = 4096$, distributed across a process grid $24 \times 32$, generating 768 total processes. The GPUs used had 132 streaming multiprocessors (SMs) based on the SM 90 architecture and operated at a peak clock frequency of 1980 MHz.

The observed maximum performance (Rmax) reached 339.86 PFLOPS overall, with each GPU contributing approximately 442.52 TFLOPS. When isolating the LU factorization phase, the performance increased to 539.19 PFLOPS, or 702.07 TFLOPS per GPU. The benchmark was executed in Sloppy FP8 mode (sloppy type = 1), and the calculated solution successfully passed the numerical validation check with a residual of $5.01 \times 10^{-5}$, well below the threshold of $1.6 \times 10^{1}$. These results highlight the strong suitability of the system for AI and mixed precision scientific computing workloads.

\begin{table}[htbp]
\centering
\caption{HPL-MxP Benchmark Summary}
\begin{tabular}{ll}
\toprule
\textbf{Item} & \textbf{Value} \\
\midrule
Benchmark version & HPL-MxP-NVIDIA 25.4.0 \\
Matrix size \(N\) & 2,989,056 \\
Block size \(NB\) & 4096 \\
Process grid (P × Q) & 24 × 32 \\
Total processes & 768 \\
Peak clock frequency & 1980 MHz \\
GPU SM version & SM 90 \\
GPU SM count & 132 \\
Observed Rmax & 3.3986e+08 GFLOPS \\
Rmax per GPU & 442520.81 GFLOPS \\
LU-only  &  5.3919e+08 GFLOPS \\
LU-only per GPU & 702,074.99 GFLOPS \\
Precision mode & Sloppy FP8 (sloppy-type = 1) \\
Validation result & PASSED (5.01e-05 < 1.6e+01) \\
\bottomrule
\end{tabular}
\label{tab:hpl_mxp_summary}
\end{table}

\paragraph{IO500}
IO500 is a benchmark suite designed to evaluate HPC storage subsystems. Unlike HPL, which measures FLOPS, IO500 assesses I/O performance through IOR and mdtest, capturing both bandwidth and metadata operations. These results are combined using a geometric mean to produce an overall score. Updated semiannually alongside the SC and ISC conferences, IO500 has become a standard metric for comparing storage capabilities in scientific computing and AI applications~\cite{kunkel:2016:establishing}.

Table~\ref{tab:io500_comparison} compares the IO500 benchmark results between a 10-node configuration and a 96-node configuration. Interestingly, while the 10-node setup outperformed the 96-node configuration in some bandwidth-heavy workloads—such as \texttt{ior-easy-write} (262.91 GiB/s vs. 198.80 GiB/s) and \texttt{ior-easy-read} (365.71 GiB/s vs. 305.86 GiB/s)—the 96-node system showed superior performance in metadata-intensive operations. This is particularly evident in tests such as \texttt{mdtest-easy-stat} (463.13 kIOPS vs. 358.75 kIOPS), \texttt{mdtest-hard-stat} (408.13 kIOPS vs. 262.43 kIOPS), and \texttt{find} (2637.17 kIOPS vs. 1976.05 kIOPS).
The overall IO500 score for the 96-node system reached 214.09, surpassing the 181.91 score of the 10-node system. This improvement is largely attributed to the scalability of metadata operations, where increased node count provides more parallelism. However, the fact that bandwidth scores were relatively close (139.80 GiB/s for 96 nodes vs. 133.03 GiB/s for 10 nodes) suggests that I/O throughput may have reached a saturation point or was limited by backend storage performance.

In summary, the comparison highlights that scaling out to more nodes can significantly benefit metadata-heavy workloads, while the gains in pure bandwidth may be more modest, depending on system architecture and workload characteristics.

\begin{table}[htbp]
\centering
\caption{Comparison of IO500 Results: 10 Nodes vs 96 Nodes}
\begin{tabular}{lcc}
\toprule
\textbf{Benchmark} & \textbf{10 Nodes} & \textbf{96 Nodes} \\
\midrule
ior-easy-write (GiB/s) & 262.91 (340.96 s) & 198.80 (355.68 s) \\
mdtest-easy-write (kIOPS) & 204.44 (347.00 s) & 256.64 (363.81 s) \\
ior-hard-write (GiB/s) & 15.84 (354.60 s) & 24.61 (491.75 s) \\
mdtest-hard-write (kIOPS) & 120.84 (340.05 s) & 151.59 (332.84 s) \\
find (kIOPS) & 1976.05 (56.39 s) & 2637.17 (54.01 s) \\
ior-easy-read (GiB/s) & 365.71 (245.15 s) & 305.86 (231.13 s) \\
mdtest-easy-stat (kIOPS) & 358.75 (197.40 s) & 463.13 (200.14 s) \\
ior-hard-read (GiB/s) & 205.64 (31.23 s) & 255.31 (48.82 s) \\
mdtest-hard-stat (kIOPS) & 262.43 (157.19 s) & 408.13 (124.54 s) \\
mdtest-easy-delete (kIOPS) & 168.19 (422.12 s) & 198.91 (468.91 s) \\
mdtest-hard-read (kIOPS) & 205.39 (200.53 s) & 310.87 (162.97 s) \\
mdtest-hard-delete (kIOPS) & 92.29 (445.91 s) & 111.28 (453.80 s) \\
\midrule
\textbf{Bandwidth Score (GiB/s)} & \textbf{133.03} & \textbf{139.80} \\
\textbf{IOPS Score (kIOPS)} & \textbf{248.74} & \textbf{327.84} \\
\textbf{Total IO500 Score} & \textbf{181.91} & \textbf{214.09} \\
\bottomrule
\end{tabular}
\label{tab:io500_comparison}
\end{table}

\section{Discussion}

At ISC 2025, the SAKURAONE system achieved a significant milestone by ranking \textbf{49th on the TOP500 list} based on its High Performance Linpack (HPL) benchmark result. This achievement underscores the system's ability to deliver globally competitive performance while adhering to an open and flexible architectural philosophy.

Among systems utilizing NVIDIA GPUs, SAKURAONE distinguishes itself as the only system within the TOP100 to employ a fully open, SONiC-based Ethernet interconnect. In contrast, most high-ranking systems rely on proprietary solutions or conform to NVIDIA’s Spectrum-X reference architecture. Ethernet-based interconnects remain rare among the upper ranks of the TOP500, and nonSpectrum-X Ethernet systems are even more rare. SAKURAONE currently holds the highest global ranking among systems that use a SONiC-based Ethernet fabric.

This achievement highlights the technical viability and competitiveness of open networking technologies in high-performance computing environments. It further demonstrates the potential of SONiC and Ethernet-based architectures as scalable, vendor-neutral alternatives to traditional HPC interconnects, enabling cost-effective and customizable system designs.

The HPCG benchmark provides deeper insight into system performance under memory- and communication-bound workloads. SAKURAONE recorded an HPCG result of 0.396~PFLOPS, corresponding to approximately 0.8\% of its theoretical peak performance as measured by HPL. While this fraction may appear modest, it is within the expected range for Ethernet-connected clusters.

SAKURAONE employs an 800~GbE Ethernet interconnect, which, compared to low-latency interconnects such as InfiniBand or Cray Slingshot, can yield lower efficiency for HPCG-type workloads. Nonetheless, the observed performance reflects a well-balanced system, and further gains may be achieved through software-level optimizations, such as improved MPI tuning, NUMA locality enhancement, and memory bandwidth maximization. These observations also help identify potential bottlenecks under realistic workload conditions and inform future system-level improvements.

In the HPL-MxP benchmark, SAKURAONE achieved 0.340~EFLOPS, placing it 12th globally. This measurement was obtained using FP8 arithmetic on NVIDIA GPUs with highly optimized binaries. The use of low-precision formats, specifically FP8, in combination with Tensor Core acceleration, resulted in a tenfold speedup compared to the system’s FP64-based HPL result of 0.0340~EFLOPS. 

This substantial gain demonstrates the architectural advantages of GPU acceleration for matrix-intensive operations. Although FP8 involves reduced numerical precision, it is highly effective for dense matrix-matrix multiplications, the core operation of the HPL-MxP benchmark. These findings confirm that SAKURAONE’s GPU nodes are not only suited to traditional HPC applications, but also to AI-centric and mixed-precision scientific workloads, where both computational throughput and efficiency are essential.

According to the IO500 results presented at ISC25, SAKURAONE achieved a total score of 181.91 in the “10 Node Production” category, ranking 9th globally and 2nd among Japanese entries. This category emphasizes practical I/O performance under production-ready configurations, balancing bandwidth, and metadata handling.
Using 10 compute nodes and 1,280 client processes, SAKURAONE delivered a bandwidth of 133.03~GiB/s and metadata throughput of 248.74~kIOP/s. These results reflect a well-tuned storage stack capable of supporting a wide range of I/O-intensive workloads, from large-scale sequential transfers to small file operations.
The system leverages DDN’s EXAScaler file system, based on Lustre, deployed on DDN ES400NVX2-NDR200 storage servers interconnected via an 800~GbE network. This configuration demonstrated outstanding metadata and bandwidth efficiency within a compact deployment footprint.

The strong IO500 performance underscores SAKURAONE’s ability to deliver production-grade I/O scalability and efficiency, even on a moderate scale. These results further validate the system's suitability for both conventional HPC workloads and modern AI applications that demand high I/O concurrency and throughput.
These benchmarking efforts in multiple application domains, ranging from LINPACK and HPCG to HPL-MxP and IO500, have enabled a comprehensive and objective evaluation of SAKURAONE. By establishing performance rankings in each category, we provide a basis for a fair comparison with other state-of-the-art HPC and AI systems.

From a performance-per-node perspective, the observed results are consistent with expectations for a system of SAKURAONE’s scale and architecture. In particular, the system demonstrates that high-ranking performance can be achieved using an Ethernet-based interconnect made up entirely of open technologies, including the SONiC network operating system, which marks a significant deviation from conventional HPC network solutions.

Although energy consumption metrics were not included in the current evaluation due to constraints in measurement conditions, future studies will incorporate power-aware benchmarking to assess energy efficiency. This will be essential to advance the design of sustainable and energy-conscious high-performance computing infrastructures.

\section{Conclusion}

In this paper, I presented the architecture and performance evaluation of SAKURAONE, a high-performance computing system designed to meet the combined demands of traditional scientific workloads and emerging AI applications. Through comprehensive benchmarking using HPL, HPCG, HPL-MxP, and IO500, I demonstrated the system’s competitiveness across a variety of metrics including floating-point throughput, memory-bound performance, mixed-precision compute, and I/O efficiency.

The results show that SAKURAONE achieves globally recognized rankings across several benchmark suites, validating the effectiveness of its design. Notably, the system is distinguished by its use of a fully open SONiC-based Ethernet interconnect, demonstrating that non-proprietary network architectures can deliver performance on par with conventional HPC fabrics.

Although energy efficiency metrics were not included in the current evaluation due to experimental constraints, I recognize the importance of sustainability and energy-aware computing. In future work, I intend to extend this evaluation to include power consumption and performance-per-watt analysis.

I believe that the insights gained from SAKURAONE will contribute to the development of future scalable, cost-effective, and open AI platforms capable of supporting the evolving landscape of compute-intensive HPC applications.

\section*{Acknowledgement}
I am deeply grateful to Mr. Takashi Inoue of SAKURA internet Inc. for his extensive support in the design of advanced network systems and the construction of physical infrastructure.  
I would also like to express my sincere gratitude to Mr. Tomohide Hattori of Prunus Solutions Inc. for his invaluable technical support and expert guidance in system integration.  
This work was supported by the Cross-ministerial Strategic Innovation Promotion Program (SIP) on “Integrated Health Care System” Grant Number JPJ012425.

\bibliographystyle{unsrtnat}
\bibliography{references}  






\end{document}